\documentclass[]{eptcs}
\providecommand{\event}{DCM}
\providecommand{\volume}{??}
\providecommand{\anno}{2009}
\providecommand{\firstpage}{1}

\def\titlerunning{Tunable resolution}
\def\authorrunning{Russ Harmer}

\begin{document}

\title{Rule-based modelling and tunable resolution}
\author{Russ Harmer
\institute{PPS, CNRS \& Universit\'e Paris Diderot--Paris 7}}
\date{}
\maketitle

\begin{abstract}
We investigate the use of an extension of rule-based modelling for cellular signalling to create a structured space of model variants. This enables the incremental development of rule sets that start from simple mechanisms and which, by a gradual increase in agent and rule resolution, evolve into more detailed descriptions.
\end{abstract}

\newcommand{\eg}{\emph{e.g.}~}
\newcommand{\ie}{\emph{i.e.}~}
\newcommand{\cf}{\emph{cf.}~}
\newcommand{\etc}{\emph{etc.}}

\section{Rule-based modelling}

Cell signalling networks involve many binding/unbinding interactions combined with post-translational modifications (PTMs)  between a large number of primarily proteins. As is by now well-known \cite{hlavacek2006rms}, this gives rise to a combinatorial explosion in the number of possible molecular species and so in the number of reactions required to define a model. For example, a protein with ten phosphorylation sites already has, in isolation, $2^{10}$ possible states which would mean $1024$ distinct molecular species. A protein-protein interaction (PPI) depending on just one of those sites should therefore be independent of the states of the other nine; this would require $2^9$ reactions, one for each possible state of those nine sites.

This problem can be solved by using \emph{rules} that act upon \emph{partially specified} molecular species; so a rule determines an entire family of reactions. In our example, this replaces the $2^9$ reactions with a \emph{single} rule that mentions only the site of interest (and the other nine sites not at all). Rule-based modelling makes explicit the notion of \emph{agent}---and, as such, is sometimes called agent-based modelling---that is left implicit in traditional reaction-based models. In that setting, an ``agent" is implicitly a combination of molecular species just as, conversely, a rule is a combination of reactions. A further refinement in this setting is that an agent has an \emph{interface}, a set of \emph{sites} through which it can bind other agents and which can have internal \emph{states} that represent, typically, the presence or otherwise of PTMs such as phosphorylation that may affect a protein's behaviour.

In this paper, we use the rule-based language Kappa \cite{danos2007rbm} as our starting point. In Kappa, there are three principal kinds of rule: bindings, unbindings and modifications. We use binding rules to create complexes by explicitly connecting agents together via their sites; a given site on an agent can be bound to at most one other site at a time. In the concrete syntax of Kappa, the two sites involved in a bond are identified by a shared (but otherwise unimportant) number prefixed by \texttt{!}:
\begin{verbatim}
C(r),C(l) -> C(r!0),C(l!0)
\end{verbatim}
So \texttt{C(r!0),C(l!0)} is nothing more than a textual syntax for the site graph \cite{danos2008rbm} consisting of two nodes---both labelled \texttt{C}, one showing a site \texttt{r}, the other a site \texttt{l}---connected by a bond between those sites. In the natural way, unbinding rules undo previous bindings; they typically split a complex into two disjoint pieces (but not if the complex were held together by a double bond, for example). In the concrete syntax of Kappa, an internal state of a site is a text string prefixed by \texttt{\~}.
Modification rules simply change such internal states. For example, the following rule says that the site \texttt{s} of agent \texttt{S} is modified from state \texttt{u} to state \texttt{p}---but only when this site \texttt{s} is bound to the site \texttt{k} of agent \texttt{E}.
\begin{verbatim}
E(k!1), S(s~u!1) -> E(k!1), S(s~p!1)
\end{verbatim}
 
\subsection{Agent variants}

Signalling proteins are generally organized by biologists into family groups (kinases, phosphatases, small G-proteins, \etc) then smaller, more precise subgroups (tyrosine kinases, MAPK phosphatases, Ras, \etc), ending with the individual proteins (erbB2, MKP-3, H-Ras, \etc). Closely related proteins often share many of their interactions and this causes a second combinatorial explosion due to the need to enumerate many \emph{instances} of essentially the same rule, \eg if four related agents \texttt{A1-4} can each bind any of three further agents \texttt{B1-3}, we have to write each instance of this binding explicitly (so 12 rules in all).

To tackle this problem, we have recently \cite{epsilon} proposed an extension of rule-based modelling where, in addition to rules, the modeller also defines an \emph{agent hierarchy}. This specifies how new agents are defined from existing ones by small modifications of their sites: a site of the parent can be deleted, renamed or duplicated in the child. The agent hierarchy is rooted by agents declared \emph{ab initio}. The descendants of an agent are called its \emph{variants}; by default, leaves are \emph{concrete} agents and the others are \emph{generic} agents.

Rules can mention agents and sites at any level of this hierarchy; a rule is \emph{generic} unless it involves only concrete agents and sites. A generic rule generally has many concrete \emph{instantations}: a concrete agent \texttt{A} inherits a rule $r$ mentioning one of its ancestors \texttt{B} provided that no site of \texttt{B} mentioned in $r$ has been deleted in \texttt{A}; if one of \texttt{B}'s sites has been duplicated in \texttt{A}, the rule $r$ is duplicated  with one instance for each duplicate. An agent hierarchy plus a set of rules thus determines, via this \emph{compilation} procedure, a typically much larger set of concrete rules. For example, we could define a new agent \texttt{T} by duplicating the site \texttt{r} of the agent \texttt{C} above:
\begin{verbatim}
T = C[r\{r1 r2}]
\end{verbatim}
The above binding rule for \texttt{C} becomes two binding rules for \texttt{T}: one for the sites \texttt{l} and \texttt{r1}, the other for \texttt{l} and \texttt{r2}. As a consequence, whereas \texttt{C} builds up linear polymers, \texttt{T} builds (binary) branching polymers.

This framework allows a clear uniform treatment of: agents representing products of closely related genes, \eg ERK1 and ERK2; agents representing splice variants of a single gene, \eg p66, p52 and p46 Shc; and agents representing mutated forms of proteins, \eg the L858R substitution of EGFR. An agent hierarchy can be thought of as an abstract, \ie not necessarily factual, phylogenetic tree; as such, in \cite{epsilon}, agent hierarchies were restricted to be trees. In this paper, we weaken this to DAGs; hierarchies need not therefore correspond to authentic phylogenetic trees, but we will see that this affords pragmatically useful additional expressive power.

\subsection{Agent resolution}

There is a subtle tension in representing biological knowledge. On the one hand, a number of online databases (\eg UniProt, HPRD) document, in great detail, the sequences, domains, modifiable residues and interactions of proteins but do not provide any formalization of this knowledge. On the other hand, formal models rarely document proteins but instead provide an executable representation of their interactions, \eg as ODEs. Indeed, it might not even be desirable to represent all information in a model, \eg it might suffice to consider all products of the same gene (or even several related genes) as a single entity in order to illustrate a desired point.

The rule-based approach takes a first step towards resolving this issue. A rule is a formalization of a little nugget of biological knowledge, \eg `the SH2 domain of Grb2 binds the phosphorylated Tyr317 residue of p52 Shc'. As such, a rule-based model is essentially a \emph{self-documenting collection of facts} that is, moreover, directly executable \cite{danos2007ssc}. This contrasts with most reaction-based models which, while executable, disperse facts across multiple reactions as a consequence of the need to fully specify all molecular species, \cf the dispersion of agents across molecular species.

The introduction of an agent hierarchy significantly extends the documentary power of the rule-based approach. First of all, the very definition of a hierarchy already collates a significant amount of biological knowledge and can serve as useful documentation for families of proteins. More importantly, an agent hierarchy is written down once and for all and yet, in conjunction with a single generic rule set, can be used to generate multiple concrete rule sets just \emph{by varying the choice of concrete agents}. (Recall that, by default, these are the leaves of the hierarchy.)

For example, the above rule codifying Grb2's binding to p52 Shc actually applies to all three splice variants of Shc and to two distinct residues of each Shc variant, although the precise numbering of the residues is different in each. We can easily express this with a simple hierarchy of Shc agents and a single generic rule:
\begin{verbatim}
Shc(PTB,YXNX~u,SH2)
Grb2(SH3n,SH2,SH3c)

Shc(YXNX~p), Grb2(SH2) -> Shc(YXNX~p!0), Grb2(SH2!0)

p66 = Shc[YXNX\{Y349 Y427}]
p52 = Shc[YXNX\{Y239 Y317}]
p46 = Shc[YXNX\{Y194 Y272}]
\end{verbatim}
If \texttt{p66}, \texttt{p52} and \texttt{p46} are taken to be the concrete agents (as they are by default), the generic rule compiles into six concrete rules, two for each variant of Shc. However, if we specify \texttt{Shc} to be concrete, the ``generic" rule is actually concrete (it is its own compilation).

This means that a modeller can transparently shift between different levels of \emph{agent resolution}: if certain details are unimportant, they can be filtered out by designating the concrete agents to be higher up in the hierarchy; if it turns out later that those details actually are important, the preceding design choice can be reversed simply by moving the concrete fringe back down and recompiling the generic rules. Usually, such design choices are hard-wired into the details of a model and cannot just be undone without forcing wholesale changes to the model. The compilation of a generic rule set with respect to a particular choice of concrete agents therefore gives modellers the ability to \emph{tune} the level of agent resolution for a given model; indeed, a generic rule set is not one model but many---parametrized by the placement of the concrete agents in the hierarchy.

\subsection{Rule resolution}

We now turn to the related notion of \emph{rule resolution} by which we mean  the extent to which (generic) rules are restricted to only certain subfamilies of agents. For example, a highly generic rule for the docking between MAP2Ks and MAPKs would be perfectly sufficient to study typical properties of an isolated MAPK cascade:
\begin{verbatim}
MAP2K(D,AS,atp,S~u,ST~u,cat~n)
MAPK(CD,AS,atp,T~u,Y~u,cat~n)

MAP2K(D), MAPK(CD) <-> MAP2K(D!0), MAPK(CD!0)
\end{verbatim}
(Note that this docking rule is reversible and so is actually two rules: docking and undocking.)

If we then wished to turn our attention to the behaviour of particular human MAPK cascades, we might extend the hierarchy along the following lines.
\begin{verbatim}
ERKK = MAP2K
JNKK = MAP2K
p38K = MAP2K
ERK = MAPK
JNK = MAPK
p38 = MAPK

ERKK(D), ERK(CD) <-> ERKK(D!0), ERK(CD!0)
JNKK(D), JNK(CD) <-> JNKK(D!0), JNK(CD!0)
p38K(D), p38(CD) <-> p38K(D!0), p38(CD!0)
\end{verbatim}

Note that the cascade has become three parallel cascades. The original docking rule would now generate far too many concrete rules---most of which would violate the desired binding specificity of \texttt{ERKK} for \texttt{ERK}, \texttt{JNKK} for \texttt{JNK} and \texttt{p38K} for \texttt{p38}---so we have replaced it with more specific \emph{instantiations} that ``insulate" the cascades from each other. This notion of (generic) rule instantiation is perfectly well-defined, even if we replace a generic agent by another generic agent below it in the hierarchy. For example, if we further extended our hierarchy with
\begin{verbatim}
MEK1 = ERKK
MEK2 = ERKK
\end{verbatim}
the instantiation of \texttt{MAP2K} by \texttt{ERKK} and \texttt{MAPK} by \texttt{ERK} would yield the now-generic  rule
\begin{verbatim}
ERKK(D), ERK(CD) <-> ERKK(D!0), ERK(CD!0).
\end{verbatim}
Rule instantiation is thus a sort of \emph{partial} compilation of generic rules into concrete rules.
It is also analogous to rule \emph{refinement} \cite{danos2008rbm}; indeed, in many agent hierarchies, all variants of some root agent can be seen, in pure rule-based style, as instances of the same agent in different states (\eg via ``private" sites with internal states encoding the agent's ``real" name in the agent hierarchy). The process of rule instantiation then corresponds to refinement of that rule to accept only certain states (\ie certain descendants) of the agent [V.~Danos, private communication]. Of course, the special case of instantiating rules with only concrete agents corresponds exactly to the rule compilation described in \cite{epsilon} (and outlined in \S 1.1).

It is important to note that, in the previous section, the agent hierarchy was varied (in the choice of concrete agents) while the generic rules remained fixed. This permits the modeller to produce a range of rule sets that vary in their level of agent detail. In this section, however, it is the rules that are varied (by instantiation) while the hierarchy remains unchanged. Instead of tuning the agent resolution of the induced concrete rule set, this tunes the \emph{rule resolution} and leads to a more carefully generated (and typically much smaller) concrete rule set. We thus have two independent axes of tunable resolution. In the next section, we examine real examples of this coming from the interactions between MAPKs and their phosphatases and upstream activators respectively.

\paragraph{Context}
This work  belongs to a large and growing literature based on the approach, initiated by Regev \cite{Regev3}, of using process-theoretic ideas (from concurrency theory) to represent proteins and their networks of interactions. While very promising, this approach is not always straightforward to use as models rapidly grow in size and complexity, becoming unwieldy and hard to maintain and further develop; furthermore, biological knowledge is increasing all the time, so a realistic modelling framework must be sufficiently intuitive, yet flexible enough to accommodate regular updates; cf. $\!$\cite{guerriero2007atn}. The ability to tune agent and/or rule resolution addresses this by allowing for incremental development of models: instead of having to work out and express everything in one go, a first approximation can be made which will be subsequently refined by extending agent hierarchies and/or instantiating generic rules to get more precise descriptions.
\section{Tunable resolution}

\subsection{MAPK--MAPK phosphatase interaction}

MAPK phosphatases come in three basic flavours: S/T- (serine/threonine) specific, Y- (tyrosine) specific and dually specific phosphatases. We can organize this into a simple V-shaped (DAG) hierarchy:
\begin{verbatim}
MAPKSTP(AS,cat~n)
MAPKYP(AS,cat~n)

MAPKSTP(AS,cat~y), MAPK(T~p,Y) -> MAPKSTP(AS!0,cat~y), MAPK(T~p!0,Y) @ b (u)
MAPKSTP(AS!0), MAPK(T!0) -> MAPKSTP(AS), MAPK(T)
MAPKSTP(AS!0,cat~y), MAPK(T~p!0) -> MAPKSTP(AS!0,cat~y), MAPK(T~u!0)

MAPKYP(AS,cat~y), MAPK(T,Y~p) -> MAPKYP(AS!0,cat~y), MAPK(T,Y~p!0) @ b (u)
MAPKYP(AS!0), MAPK(Y!0) -> MAPKYP(AS), MAPK(Y)
MAPKYP(AS!0,cat~y), MAPK(Y~p!0) -> MAPKYP(AS!0,cat~y), MAPK(T~u!0)

MKP = MAPKSTP[+KIM]
MKP = MAPKYP[+KIM]

MKP(KIM), MAPK(CD) <-> MKP(KIM!0), MAPK(CD!0)
\end{verbatim}
Note that \texttt{MKP} has two distinct definitions; it is an \emph{alias} for its two parents. This means it inherits from both and so acquires the necessary rules to modify \texttt{MAPK} on both \texttt{T} and \texttt{Y}. It has also acquired a new site \texttt{KIM} which is used to dock with \texttt{MAPK}s. The rules are textbook examples of enzyme-substrate interaction; the only subtlety lies in their kinetic rates (\texttt{b} and \texttt{u}). In rule-based modelling, a binary rule, while usually causing a binary reaction (with rate \texttt{b}), may also provoke a \emph{unary} reaction: the two agents binding may \emph{already be bound} on other sites in which case \texttt{u} (units $s^{-1})$ is the corresponding first-order rate constant.

\subsection{Increasing resolution}

Let us now extend the hierarchy to add some additional agents and generic rules:
\begin{verbatim}
HePTP = MAPKYP
ERKP = MPK
JNKP = MPK
p38P = MPK

ERKP(KIM), ERK(CD) <-> ERKP(KIM!0), ERK(CD!0)
JNKP(KIM), JNK(CD) <-> JNKP(KIM!0), p38(CD!0)
p38P(KIM), JNK(CD) <-> p38P(KIM!0), p38(CD!0)

MKP3 = ERKP
MKP1 = ERKP
MKP1 = JNKP
MKP1 = p38P
DUSP5 = JNKP[cat~y]
DUSP5 = p38P[cat~y]

MKP3(KIM!1,cat~n), ERK(CD!1) -> MKP3(KIM!1,cat~y), ERK(CD!1)
MKP1(KIM!1,cat~n), MAPK(CD!1) -> MKP1(KIM!1,cat~y), MAPK(CD!1)
\end{verbatim}
Note that these three new docking rules are instantiations of the generic MKP/MAPK docking; note also that \texttt{HePTP} does not acquire a docking rule.

The DAG structure naturally \emph{encodes} the observed facts \cite{MKP2007} that MKPs have complex, overlapping binding specificities---but does not \emph{explain} how. Some MPKs specifically bind ERK, others specifically bind JNK/p38 and others still bind all three kinds of MAPKs.
In reality, this variety of binding specificity is brought about by there being (at least) two distinct possible binding sites on the MAPK and two or three on the MKP; a more detailed hierarchy could therefore be defined that expresses this additional detail. This would constitute a better formal expression of the actual molecular mechanism used by MKPs and MAPKs to dock, not just the functional consequences of that mechanism that the DAG encodes; but we must leave the construction of that more detailed model, and comparison with the present model, for future work.

The agent \texttt{DUSP5} adopts a new default internal state \texttt{y} for its site \texttt{cat} whereas \texttt{MKP3} and \texttt{MKP1} acquire this state only after docking, presumably via an allosteric mechanism \cite{MKP2007}. This reflects the fact that the enzyme-substrate interaction between either \texttt{MKP1} or \texttt{MKP3} and \texttt{MAPK}s only occurs as a monomolecular reaction (within an \texttt{MKP-MAPK} complex); whereas \texttt{DUSP5} can additionally engage in docking-independent enzyme-substrate interaction.

It would obviously be possible, although cumbersome, to transform this hierarchy into a tree by the usual algorithm that expands any DAG to a tree; but it would be much more interesting to attempt to build a tree hierarchy that accurate reflects the actual phylogeny of the MKP protein family. It is to be hoped that the corresponding rule set would be fairly generic since one would expect close phylogenetic relatives to share many interactions; but this, too, we must leave for future investigation.

In summary, we have navigated from a starting, ``simple" rule set that is generated by taking MKP to be a concrete agent and using the completely generic docking rule, to a ``detailed" rule set generated by taking MKP as a generic agent with several distinct concrete children and the three more specific instantiations as docking rules. Note that the agent hierarchy itself \emph{does not change}; only our choice of concrete agents with respect to which we compile. Analogously, the rule set \emph{does not change} either; it is our choice of rule instantiations that varies. In other words, these two very different rule sets actually coexist within a single model; and we can choose between them simply by picking the appropriate agent and rule resolutions and compiling down to concrete rules.

There is a sense in which we can consider the detailed rule set as the result of an \emph{evolution} of the simple rule set. A generic rule can either \emph{adapt} to the introduction of new agent variants, \eg the docking rule that is replaced by its three instantiations, or be \emph{conserved} across the new agents, \eg the enzyme-substrate rules for \texttt{MKP}s that do not take any account of the refinement of \texttt{MKP} or \texttt{MAPK} into subfamilies. With hindsight, the erbB rule set developed in \cite{epsilon} can also be seen in this way as an evolution from a single ligand-receptor interaction to a rich network of ligand families and receptors with complex, overlapping binding specificities.

\subsection{Distributive vs.~processive activation}

As a final illustration, we outline how the evolution of rule sets can even modify the ways in which rules can be applied so as to change the molecular mechanism they represent. We begin with the \texttt{MAP2K} and \texttt{MAPK} agents from \S 1.3 and add the standard enzyme-substrate rules precisely as for \texttt{MKP}.

\begin{verbatim}
MAP2K(AS,cat~y), MAPK(T~u,Y) -> MAP2K(AS!0,cat~y), MAPK(T~u!0,Y) @ b (u)
MAP2K(AS!0), MAPK(T!0) -> MAP2K(AS), MAPK(T)
MAP2K(atp,AS!0,cat~y), MAPK(T~u!0) -> MAP2K(atp,AS!0,cat~y), MAPK(T~p!0)

MAP2K(AS,cat~y), MAPK(T,Y~u) -> MAP2K(AS!0,cat~y), MAPK(T,Y~u!0) @ b (u)
MAP2K(AS!0), MAPK(Y!0) -> MAP2K(AS), MAPK(Y)
MAP2K(atp,AS!0,cat~y), MAPK(Y~u!0) -> MAP2K(atp,AS!0,cat~y), MAPK(Y~p!0)
\end{verbatim}
In isolation, these rules implement what biochemists call a \emph{distributive} dual phosphorylation mechanism: the enzyme must unbind its substrate after each modification, so two independent enzyme-substrate interactions are needed for a doubly-phosphorylated substrate to come about. This mechanism is often found in real MAPK cascades, particularly in (relatively speaking) simple organisms such as xenopus oocytes \cite{ferrell1997msd}, and gives rise to a sharply ultrasensitive dose-response curve.

If we now reincorporate the simple docking rule from \S 1.3,
\begin{verbatim}
MAP2K(D), MAPK(CD) <-> MAP2K(D!0), MAPK(CD!0)
\end{verbatim}
the above enzyme-substrate interaction rules can be applied to either docked or undocked agents. In particular, it is possible for a \texttt{MAP2K} to dock with a \texttt{MAPK} and modify \emph{both} its sites before undocking. This implements what biochemists call a \emph{processive} dual phosphorylation. Such a mechanism typically produces a less sharp does-response curve as, unlike in the distributive case, doubly-phosphorylated substrate can start to appear immediately.

Note that, as in the \texttt{MKP} example, we have left the enzyme-substrate rules unchanged, \ie we consider them to be conserved---and hence promiscuous---across all variants of \texttt{MAP2K} and \texttt{MAPK}. Alternatively, we could have imposed binding specificity at the level of these rules by instantiating them with the more specific agents (\texttt{ERKK}, \texttt{JNKK} and \texttt{p38K} from \S 1.3). However, it seems more plausible from an evolutionary point of view that fine-tuned binding specificity arises from the appearance of docking sites not directly implicated in the enzyme mechanism---as ``unlucky" mutations in this mechanism might risk destroying the protein's catalytic activity. In any case, the choices we make of when to instantiate (\ie adapt) and when not to (\ie remain conserved) determine a rudimentary formal description (realistic or not) of the evolution of a rule set.

\section{Discussion}

We have presented a framework for rule-based modelling where multiple (concrete) models can be extracted from a single generic rule set and associated agent hierarchy. As such, it allows modellers to navigate around a space of related (concrete) models, selecting appropriate levels of detail according to the question in hand.

In particular, this allows for the incremental development of models where one starts from a rather simple-minded mechanism and gradually refines this into a detailed model by increasing agent and/or rule resolution, \eg the MAPK model described in this paper (MAP2K, MAPK and MKP). As a bonus, the use of generic rules also greatly speeds up the process of model construction, not least because as one is no longer drowning in a sea of concrete rules.

More conceptually, we view this process of ``unveiling" a detailed rule set from a modest one---by gradual increasing agent and rule resolution---as the first steps towards a formalization of what it means to ``evolve" a rule set. For example, the MAP2K/MAPK rule set starts as a rudimentary enzyme-substrate interaction in a single cascade activated by a distributive mechanism, which evolves into multiple cascades activated by a docking-mediated processive mechanism.

The next step, as briefly mentioned above, would be to produce a more phylogenetically-accurate agent hierarchy, accompanied by a more precise generic rule set, in an attempt to explain how, for example, MKP/MAPK docking specificity is implemented at the molecular level. In order to do this, we might need to move to a deeper level of representation of proteins: rather than representing a protein as a single agent, instead consider individual domains as agents and proteins as unbreakable complexes, connected by a backbone. This representation would enable separate hierarchies for domains, not just proteins, which would afford far more subtle possibilities for protein construction by recombination of elements of these hierarchies.

\paragraph{Acknowledgements}
My thanks to Vincent Danos, J\'er\^ome Feret, Walter Fontana and Jean Krivine for numerous discussions related to this work.

\bibliographystyle{unsrt}
\bibliography{bio}

\end{document}